\documentstyle{article}

\begin{document}
\title{Probing the electron charge distribution via Kapitza-Dirac diffraction}
\date{}
\author{Pedro Sancho \\ Centro de L\'aseres Pulsados CLPU \\ Parque Cient\'{\i}fico, 37085 Villamayor, Salamanca, Spain}
\maketitle
\begin{abstract}
We analyze the diffraction of elementary systems as the electron by
light gratings when they are described by charge distributions
instead of the usual point-like form. The treatment of the problem
is based on the introduction, in analogy with atomic polarizability,
of state-dependent non-permanent multi-pole moments for the charge.
The diffraction patterns can provide bounds on these  moments. With
this approach we can experimentally explore some aspects of the
interpretation of the wave picture of single charges.
\end{abstract}

\section{Introduction}

The question of the elementary character of seemingly structureless
particles as  the electron frequently emerges in the physical
literature. These analysis mainly focus on three interconnected
aspects: the existence of substructures \cite{Eic,Bou}, the presence
of dipole or higher moments \cite{Pos,Hud}, and the actual size of
the electron (as opposed to a point-like object) \cite{Gab,Deh}.
These works are based on a large variety of experimental techniques
such as Thomson and Compton scattering, Penning traps, high energy
electron-positron collisions or precision measurements of the energy
states of YbF molecules.

We consider in this paper a particular aspect of the problem,
closely related to the wave-particle duality. Quantum systems can
behave as particles (point-like) or waves (spatially extended). The
typical examples of the last case are diffraction experiments. We
can ask for the behavior of a charge in this framework: does it also
show spatially extended characteristics? Equivalently, does it
possess multi-pole moments different from zero in these
circumstances? We address some aspects of the question by invoking
techniques of optical (Kapitza-Dirac effect \cite{KD,BaC}) and atomic
(non-permanent moments \cite{Sch}) physics.

Comparing the diffraction patterns of point-like and spatially
extended charges we can study the differences between the two types
of charge distributions. In order to carry out the comparison one
must derive the expected patterns in both cases. The laser
interaction with a point-like charge is given by the ponderomotive
potential \cite{Fed}. We must evaluate the laser interaction with a
charge distribution. When the characteristic length of the charge
distribution is smaller than the wavelength of the driving laser we
can approach the problem via a multi-pole expansion of the
distribution. Using this expansion it is simple to derive the
diffraction patterns for this case in the approximation of initial
plane wave states. This simple but illustrative example shows that
the patterns differ in both cases. Determining these detection
probabilities we can introduce bounds on the values of the
multi-pole moments. The advantage of the Kapitza-Dirac effect is
that the experiment with electrons has been carried out in
\cite{BaN}. Minor modifications of the arrangement, introduced to
use adequate laser wavelengths, would allow us to confront
theoretical results with actual data.

From the beginning it must be stressed that the moments we study in
the paper would not be intrinsic or permanent to the electron.
Permanent moments would be present under any circumstances. There
have been experiments with YbF molecules providing precision bounds
on their possible values \cite{Hud}. In contrast, we are interested
into moments that would only manifest when the charge is forced to
behave as an extended object, that is, non-permanent moments. There
is a clear analogy of this proposal with atomic polarizability,
where an external field can induce non-intrinsic moments in the atom
\cite{Sch}.

\section{Ponderomotive potential for distributions}

Our first task is to determine the form of the light-matter
interaction for charge distributions. We must know the form of the
interaction potential to be introduced into the Schr\"odinger
equation. As in the point-like case we assume that this potential
has the classical form. Then we must derive the form of the
interaction between the charge distribution and the light field in
the classical electromagnetic formalism. We expect, by similitude
with the point-like case, that the Schr\"odinger equation with the
interaction potential of classical charge distributions will
correctly describe the quantum behavior of extended charges. In this
section, from now on, all our considerations will be from a
classical nature. In particular, the charge density associated with
the elementary charge is classical.

The interaction of a point-like electron with a laser is described by the ponderomotive force,
\begin{equation}
{\bf F}_P=-\frac{e^2}{4m\omega _L^2}\nabla {\bf E}^2({\bf r})
\end{equation}
which can be expressed in terms of a potential, ${\bf F}_P =-e \nabla V_P$, denoted as the ponderomotive potential:
\begin{equation}
V_P ({\bf r})= \frac{e}{4m\omega _L^2}{\bf E}^2({\bf r})
\end{equation}

with ${\bf E}$ the electric field of the laser and $\omega _L$ its
frequency (the temporal dependence is averaged over the
laser period). The energy of the charge in this potential is $eV_P$.

In the particular case of a standing light wave with
spatial and temporal dependence $\cos k_Lx \cos \omega _L t$, with
$k_L$ the light wavelength and $x$ the coordinate in the laser propagation direction, it can be written as
\begin{equation}
V_P(x)=\frac{e{\bf E}_0^2}{4m\omega _L^2} \cos ^2k_Lx
\end{equation}
We consider now how an electron behaves in that potential when we
assume that the charge is a distribution instead of a point-like
object. The energy of a charge density distribution, $\rho ({\bf r})$, in a
ponderomotive potential is
\begin{equation}
U_P = \int \rho ({\bf r})V_P({\bf r})d^3 {\bf r}
\end{equation}
This expression gives the total energy of the charge in the
ponderomotive potential. There are two regimes. The first one refers
to situations where the relation between the typical scales of the
charge distribution and the wavelength of the laser allows for a
multi-pole approach to the problem. When this is not possible, the
second regime, we must resort to other techniques. In this paper we
shall restrict our considerations to the first case.

When the potential varies smoothly in the region where $\rho ({\bf
r})$ is defined we can expand $V_P$ around a point taken as the
origin \cite{Jac}
\begin{equation}
V_P({\bf r})=V_P(0)+{\bf r} \cdot (\nabla V_P)(0)+\frac{1}{2}\sum _{i,j} r_ir_j \frac{\partial ^2 V_P}{\partial r_i \partial r_j} (0) + \cdots
\end{equation}
Introducing this expression into the energy equation and using the
relation $\int \rho ({\bf r})d^3 {\bf r}=e$, we have
\begin{equation}
U_P=eV_P(0)+ {\bf D}\cdot (\nabla V_P)(0) + \frac{1}{2}\sum _{i,j} Q_{ij} \frac{\partial ^2 V_P}{\partial r_i \partial r_j} (0) + \cdots
\end{equation}
with the usual dipole ${\bf D}=\int {\bf r}\rho ({\bf r})d^3{\bf r}$
and quadrupole moments $Q_{ij}=\int r_ir_j \rho ({\bf r})d^3 {\bf
r}$. It is important to remark that in our problem the moments do
not refer to a set of charges but to a single charge with a spatial
distribution.

In order to the multi-mode expansion be useful we expect the contribution of
the terms to decrease when its order increases. When this is so, only
the zero, dipole and quadrupole moments are relevant and we can
neglect the rest. We analyze this point later.

\section{Diffraction patterns for distributions}

Once derived the interaction potential for point-like and extended
charges we can study the quantum diffraction patterns in both cases.
We consider a situation that can be solved analytically, that where
the initial state of the electron can be described by the
approximation of a plane wave. Of course, to use plane waves is an
oversimplification in realistic problems. However, it provides a
simple example that illustrates the differences between both
approaches.

In this approximation the solution for a point-like charge is well-known
\cite{BaC}. The problem can be taken as an one-dimensional one,
where we only must care about the transversal variables, that is,
those in the direction of propagation of the light. The diffractive
regime is reached when the potential is much larger than the recoil
shift ($\epsilon =\hbar ^2 k_L^2/2m$). This condition is equivalent
to neglect the free term of the Hamiltonian when compared to the
interaction term (Raman-Nath approximation) \cite{BaC}. In this
regime the initial plane wave state of the electrons, $\psi (0)=
e^{ik_0x}$ with $k_0$ the initial wavelength, evolves as
\begin{equation}
\psi (t)=e^{ieV_P(x)t/\hbar}\psi (0)=e^{ieV_0t/2\hbar } \sum _{n=-\infty}^{\infty}i^n J_n \left( \frac{eV_0t}{2\hbar} \right)e^{i(2nk_L+k_0)x}
\end{equation}
where we have introduced the notation $V_0=e{\bf E}^2/4m\omega
_L^2$ and we have used the expression $e^{i\xi \cos \varphi}=\sum
_{n=-\infty}^{\infty} i^n J_n (\xi)\exp(in\varphi)$, with $J_n$ the $n$-th order Bessel's function. The wavelength
of the electron changes by even multiples of $k_L$. The probability
of detecting the particle in the $n$-th diffraction order is given
by $|J_n|^2$.

We move now to the case of charge distributions. The energy $U_P$ in a standing wave reads
\begin{equation}
U_P(x)= eV_0 \cos ^2 k_Lx -V_0 Dk_L \sin 2k_Lx - V_0Qk_L^2 \cos
2k_Lx + \cdots
\label{eq:bue}
\end{equation}
If we assume, as in the standard case, that $U_P \gg \epsilon $ the evolution of the initial state is
\begin{eqnarray}
\psi (t)=e^{iU_P(x)t/\hbar}\psi (0)=e^{ieV_0t/2\hbar } \sum
_{n=-\infty}^{\infty}\sum _{m=-\infty}^{\infty} i^n \times \nonumber
\\ J_n \left( \frac{V_0t}{\hbar}\left( \frac{e}{2}-Qk_L^2 \right)
\right) J_m \left( -\frac{V_0Dk_Lt}{\hbar}
\right)e^{i(2(n+m)k_L+k_0)x}
\end{eqnarray}
In the derivation we have used the relation $e^{i\xi \sin
\varphi}=\sum _{n=-\infty}^{\infty} J_n (\xi)\exp(in\varphi)$.

The final state is a sum over all the states whose wavelength
differs from $k_0$ by even multiples of $k_L$. Both expressions, for
point-like and extended charges, are sums in a basis of plane waves
but with different coefficients. Then both diffraction patterns are
different. The simplest form to quantify these differences is to
determine the weights of each mode. This is equivalent to evaluate
the probability of diffraction with a given value of momentum
change. For instance, for the zero order (no diffraction) in the
first case we have a probability of detection
$|J_0(eV_0t/2\hbar)|^2$, whereas for a distribution charge it is
$|J_0(V_0t(e/2-k_L^2Q)/\hbar)J_0(V_0k_LDt/\hbar)|^2$. Both
expressions are clearly different. Varying the free parameters of
the problem $V_0$ (intensity of the laser) and $t$ (interaction time
between laser and electron), we could fit the last equation to the
experimental data and to determine $D$ and $Q$.

As signaled before, the approach followed here only makes sense when
the lower terms are the relevant ones. This condition can be easily
obtained with the above expressions. As $V_0$ is common to all the
orders, and the trigonometric functions are bound (their absolute
values are equal or smaller than one) the relative intensity of the
terms is given by $1,Dk_L/e,Qk_L^2/e,\cdots Q_mk_L^m/e, \cdots $. Then we
must have $Dk_L<e$, $Qk_L^2 <e$,... and $Q_m < Q_n k_L^{n-m}, m>n$.

In actual experiments, such as that reported in \cite{BaN}, the
plane wave approximation is clearly an oversimplification. In order
to compare the expected results with those of realistic experiments
we must carry out a numerical simulation of the Schr\"odinger
equation for the arrangement with the interaction potential of Eq.
(\ref{eq:bue}) and the actual state in the experiment. Varying $D$
and $Q$ and comparing with the experimental results we could discard
ranges of values incompatible with the experimental data.

We must also evaluate the order of magnitude of the characteristic
lengths that can be actually explored with the multi-pole expansion.
We denote by ${\cal L}$ the characteristic length. The standard
condition to observe an object with light is ${\cal L}\approx
\lambda _L$, that is, the laser wavelength must be of the order of
the characteristic length. At present, the shorter laser wavelengths
for which the Kapitza-Dirac effect seems to be experimentally
accessible lie on the X-ray domain \cite{San}. Beyond that domain,
in the actual state of the art, there are not coherent light sources
of enough intensity. Thus, with present day technology we could
explore via Kapitza-Dirac diffraction multi-pole structures of the
order of ${\cal L}\approx 10^{-10}m$.

Finally, we want to remark again on the concept of non-permanent
moments. A charge distribution can deviate from the point-like form
in two different ways. The charge can have an intrinsic structure
independently of its behavior or state. We say that it is a
permanent contribution and we can speak of permanent moments. On the
other hand, if the charge distribution can change with the state the
electron it can acquire additional moments. They are dependent on
the state of the system and are denoted non-permanent moments. The
situation resembles that in atomic physics, where the moments of an
atom are different when it is placed in an external electric field
due to polarizability effects \cite{Sch}. In both cases we need an
external process forcing the electron (diffraction grating) or the
atom (electric field) to a behavior where the non-permanent
contributions manifest. In \cite{San} non-permanent atomic moments
are briefly discussed in the context of Kapitza-Dirac diffraction.
The moments used here are non-permanent, they are only present when
the electron diffracts.

\section{Extended properties and the wave picture}

In the second part of the paper we suggest that the above scheme can
be used to study an interpretational question, that related to the
physical meaning of the wave picture. In diffraction experiments we
associate a wave picture with the system. Does this wave picture
describe objective extended physical properties or it is only
related to the statistical nature of the wave function reflected in
the possibility of detecting the electron at different places after
the diffraction? Clearly, one of these objective properties would be
the existence of (non-permanent) multi-pole moments different from
zero.

The natural framework for this discussion is the wave-particle
duality, where one associates exclusive wave or particle pictures
with the system. In the standard complementarity formalism the wave
picture is interpreted in a statistical sense. We do not deal with a
physical wave. The spatially extended properties of the system, for
instance the detection of the particle at different locations in
different repetitions of the experiment, only correspond to a
statistical feature of the mathematical description. In contrast,
other authors have suggested that a more physical explanation is
possible. In this alternative approach the system would posses
objective (and, in principle, testable) extended physical
properties.

In our case these properties should be associated with the charge of the
electron, the relevant element in the light-matter interaction.
First of all, it is evident that if we associate a wave picture with
the electron during the diffraction process then, because of the
unicity of the system, we must also use a wave picture for the
charge. In a physical (non-statistical) interpretation the charge would be an
extended object and, in consequence, it could be described invoking
a multi-pole formalism. These moments would not be present in the
particle picture and, consequently, must be considered from a
non-permanent nature. In contrast, in the statistical interpretation
the charge does not have extended properties, which only manifest in
the detection at different places of the point-like electron.

From a more technical point of view, if the charge is an extended object
its wave function must be calculated from an evolution equation
taking into account this extended aspect. As discussed in Sect. 2 we
must use the Schr\"odinger equation with the $U_P$ potential. The
solution of the equation will include the contributions of the
hypothetical non-permanent multi-pole moments. In contrast, in
the statistical framework one uses the evolution equation with the
point-like potential.

We summarize the above considerations. The diffraction patterns
provide an experimental method to discriminate between the physical
and statistical interpretations of the wave picture. If one can really
ascribe objective extended physical properties to the wave, it is
natural to identify them with the existence of non-permanent moments different
from zero, which lead to testable deviations from the statistical
interpretation. If no definitive deviation is obtained in the
experiments, these tests will at least provide bounds on the values
of the non-permanent moments, empirically constraining physically objective
interpretations.

\section{Discussion}

We have studied Kapitza-Dirac diffraction of electrons when the
charge is described by a distribution instead of the usual
point-like idealization. Our fundamental technical assumption is
that the evolution of the system is ruled by the Schr\"odinger
equation, but with the laser-electron interaction potential given by
$U_P$, the generalization of the ponderomotive potential to charge
distributions. With this assumption it is simple to derive the
diffraction patterns when the size of the distribution is small in
comparison with the laser wavelength. Here, we have only considered
the oversimplified case of electrons initially in plane wave states.
For more realistic initial electron states one must resort to
numerical simulations. By comparison to repetitions of the
experiment \cite{BaN} with adequate laser wavelengths and electron
states, we could infer bounds on the values of the multi-pole
moments. When the conditions for a multi-pole expansion do not hold
we should consider other types of techniques.

Our proposal has two possible applications. On the one hand, it
provides a new method to study deviations from the point-like form
in elementary systems. On the other hand, it introduces a novel tool
to analyze some aspects of the interpretation of the wave picture.

With respect to the first point we must compare our proposal with
other approaches raising the same question. As signaled in the
Introduction very stringent tests of the electron size have been
conducted using Penning traps \cite{Gab,Deh}. In these experiments
the magnetic moment of the electron is determined with very high
accuracy, agreeing extremely well with the values predicted by
Quantum Electrodynamics. Introducing models of finite size electrons
the authors derived upper limits for the electron size of
respectively $10^{-20}m$ and $10^{-22}m$. Our method provides an
alternative and independent way to estimate limits on the effective
electron size. In a related context the permanent dipole moment of
the electron has been analyzed both theoretically and
experimentally. In the Standard Model a tiny electric dipole moment,
$d_e<e \times 10^{-40}m$, is associated with the electron. This
value, although extremely small, could be used to distinguish
between the standard theory and some proposed extensions \cite{Pos}.
Recent, high accuracy measurements provide an upper limit for $d_e$
of $e \times 10.5 \times 10^{-30}m$ \cite{Hud}. The dipole moments
considered in these experiment are from the permanent type because
the electrons in Penning traps or YbF molecules are not forced to a
diffraction-like behavior. In addition, it must be noted that the
measurements in \cite{Hud} are carried out in electrons in bound
states. One must consider the possibility that dipole moments can
differ for bound and non-bound states. Clearly, the values obtained
in \cite{Hud} or predicted by the Standard Model would provide a
completely negligible contribution of the dipole term to the
modifications of the diffraction pattern. All the hypothetical
effects that could be observed in a diffraction experiment should be
attributed to non-permanent contributions.

In relation to the second point our proposal could be on the basis
of a test discriminating between statistical and physically
objective interpretations of the wave aspect of quantum systems. At least, it
can provide quantitative bounds on the viability on the second type
of interpretation. The non-permanent moments play a fundamental role
in this discussion. The polarizability effects in atomic physics
suggest an interesting analogy with these moments. Note, however,
that this analogy is only partial because in the atomic case there
is an underlying charge structure that is not present in the
electron. The non-permanent moments of the electron would be
state-dependent and could not be associated with any substructure.

To end the paper we want signal some similarities between the ideas
here presented and the work \cite{Kei}, where the radiative
properties of electrons interacting with a laser were studied as a
function of its charge distribution. These authors concluded that
during the emission process the electron cannot be treated as an
extended charge but as a point-like emitter, even when the spread of
the electron wavepacket is comparable to the wavelength of the
driving laser. The resemblances to our approach are evident.
However, there is a fundamental difference between them, in
\cite{Kei} the electron is not explicitly forced to behave in a
wave-like way. Moreover, diffraction is a smooth process whereas
emission corresponds to a sharp evolution.


\begin{thebibliography}{99}
\bibitem{Eic} Eichten E J, Lane K D and Peskin M E, 1983 Phys. Rev. Lett. {\bf 50} 811
\bibitem{Bou} Bourilkov D, 2001 Phys. Rev. D {\bf 64} 071701R
\bibitem{Pos} Pospelov M and Ritz A, 2005 Ann. Phys. {\bf 318} 119
\bibitem{Hud} Hudson J J, Kara D M, Smallman I J, Saver B E, Tarbutt M R and  Hinds E A, 2011 Nature {\bf 473} 493
\bibitem{Gab} Gabrielse G, Hanneke D, Kinoshita T, Nio M and Odom B, 2006 Phys. Rev. Lett. {\bf 97} 030802
\bibitem{Deh} Dehmelt H, 1998 Phys. Scr. T22 102
\bibitem{KD} Kapitza P L and Dirac P A M, 1933 Proc. Cambridge Philos. Soc. {\bf 29} 297
\bibitem{BaC} Batelaan H, 2000 Contemp. Phys. {\bf 41} 369
\bibitem{Sch} Schwerdtfeger P, 2006 Electric Fields (London: World Scientific)
\bibitem{Fed} Fedorov M V, 1991 Interaction of Intense Laser Light with Free Electrons (Chur: Harwood Academic Publishers)
\bibitem{BaN} Freimund D L, Aflatooni K and Batelaan H, 2001 Nature {\bf 413} 142
\bibitem{Jac} Jackson J D, 1975 Classical Electrodynamics 2nd Ed. (New York: Wiley)
\bibitem{San} Sancho P, 2013 Eur. Phys. J. D {\bf 67} 114
\bibitem{Kei} Peatross J, M\"uller C, Hatsagortsyan K Z and Keitel C H, 2008  Phys. Rev. Lett. {\bf 100} 153601
\end{thebibliography}
\end{document}